\documentclass[reprint,aip,pre]{revtex4-1}

\usepackage{amsmath}
\usepackage{amssymb}
\usepackage{graphicx}
\usepackage{color}

\begin{document}

\title{Exceeding the leading spike intensity and fluence limits in backward Raman amplifiers}
\author{V.~M.~Malkin}
\affiliation{Department of Astrophysical Sciences, Princeton University, Princeton, NJ USA 08540}
\author{Z.~Toroker} 
\affiliation{Department of Electrical Engineering, Technion Israel Institute of Technology, Haifa 32000, Israel}
\author{N.~J.~Fisch}
\affiliation{Department of Astrophysical Sciences, Princeton University, Princeton, NJ USA 08540},

\date{\today}

\begin{abstract}
The leading amplified spike in backward Raman amplifiers can reach nearly relativistic intensities before the saturation by the relativistic electron nonlinearity, which sets an upper limit to the largest achievable leading spike intensity. It is shown here that this limit can be substantially exceeded by the   initially sub-dominant spikes, which surprisingly outgrow the leading spike after its nonlinear saturation. Furthermore, an initially negligible group velocity of the amplified pulse in strongly under-critical plasma appears to be capable of delaying the filamentation instability in the nonlinear saturation regime. This enables further amplification of the pulse to even larger output fluences.
\end{abstract}


\maketitle
\section{Introduction}
The backward Raman amplification (BRA) of laser pulses in plasmas \cite{Malkin_99_PRL} is potentially capable of producing laser powers about a million times larger than the chirped pulse amplification (CPA) \cite{Mourou98} at the same wavelengths within the same size devices \cite{Malkin_00_POP,Malkin_14-EPJST}. The BRA advantage can be even much greater at laser wavelengths much shorter than 1/3 micron, where material gratings used by CPA cannot operate.
The possibility of reaching nearly relativistic unfocused intensities in backward Raman amplifiers  has  been in principle demonstrated experimentally as well \cite{Ping_04_PRL,Balakin_04_JETPL,Cheng_05_PRL,Ren_08_POP, Kirkwood_07_POP, Pai_08_PRL,Jaro_12_NJP}. 

The major physical processes that may affect BRA include the amplified pulse filamentation and detuning due to the relativistic electron nonlinearity \cite{Malkin_99_PRL,Fraiman_02_POP,Malkin_07_PRL,2012-dispersion,Malkin_14-EPJST,PoP-2014-Lehmann}, parasitic Raman scattering of the pump and amplified pulses by plasma noise \cite{Malkin_99_PRL,Malkin_00_PRL,Malkin_00_POP,Tsidulko_00_PRL,Solodov_04_PRE,Malkin_14-EPJST}, generation of superluminous precursors of the amplified pulse \cite{Tsidulko_02_PRL}, pulse scattering by plasma density inhomogeneities \cite{Solodov_dens}, pulse depletion and plasma heating through inverse bremsstrahlung \cite{Malkin_07_PRE,Malkin_09_PRE,Malkin_10_POP,2011-Balakin}, the resonant Langmuir wave Landau damping \cite{PRL-2005-Hur,Malkin_07_PRE,PoP-2009-Yampolsky,Malkin_10_POP,PoP-2011-Yampolsky,PoP-2012-Strozzi,IEEE-2014-Wu,NatCom-2014-Depierreux} and breaking \cite{Malkin_99_PRL,Malkin_00_POP,Yampolsky_08_POP,Trines_11,Malkin_14-EPJST}. 
Most of these deleterious processes can be mitigated by appropriate preparation of laser pulses and plasmas, choosing parameter ranges and selective detuning of the Raman resonance.
Ultimately, the output intensity limit appears to be imposed primarily by the relativistic electron nonlinearity (REN), causing saturation of the dominant leading spike growth. 
The major goal of this paper is to explore the possibility of extending BRA beyond this  theoretical limit for the largest achievable unfocused intensity and fluence of the output pulses.

\section{Basic Equations}
To proceed, note that the transverse filamentation instability associated with the REN can be delayed, say, by using pulses highly uniform in the transverse directions, or by transverse phase mixing.  
Then, to assess the largest output intensity, a one-dimensional model may be adequate. 
The one-dimensional equations for the resonant 3-wave interaction, together with the lowest order relativistic nonlinearity and group velocity dispersion terms, can be put in the form \cite{Malkin_07_PRL}:
\begin{eqnarray}\displaystyle
&a_t+ c_a a_z=V_3 fb~,\;\;\;\;  f_t=-V_3 ab^*, \label{1} \\
&b_t-c_b b_z=-V_3 af^* +\imath R|b|^2b  -\imath \kappa b_{tt} \, . \label{2}
\end{eqnarray}
Here $a$, $b$ and $f$ are envelopes of the pump pulse, counter-propagating shorter pumped pulse and resonant Langmuir wave, respectively; subscripts $t$ and $z$ signify time and space derivatives; $c_a$ and $c_b$ are group velocities of the pump and amplified pulses;  $V_3$ is the 3-wave coupling constant (real for appropriately defined wave envelopes), $R$ is the coefficient of nonlinear frequency shift due to the relativistic electron nonlinearity, $\kappa=c_b'/2c_b$ is the group velocity dispersion coefficient ($c_b'$ is the derivative of the amplified pulse group velocity over the frequency).  

The group velocities $c_a$ and $c_b$ are expressed in  terms of the respective laser frequencies $\omega_a$ and $\omega_b$ as follows:
\begin{equation}\displaystyle
c_a=c\sqrt{1-\frac{\omega_e^2}{\omega_a^2}}, \;\quad 
c_b=c\sqrt{1-\frac{\omega_e^2}{\omega_b^2}}\, ,
\label{e3}
\end{equation}
where $c$ is the speed of light in vacuum,
\begin{equation}\displaystyle
\omega_e=\sqrt{\frac{4\pi n_e e^2} {m_e}}
\label{e4}
\end{equation}
is the electron plasma frequency, $n_e$ is the electron plasma concentration, $m_e$ is the electron rest mass and $-e$ is the electron charge,
so that
\begin{equation}\displaystyle
2\kappa=\frac{c_b'}{c_b} = \frac{\omega_e^2}{\omega_b(\omega_b^2-\omega_e^2)}=
\frac{\omega_e^2\, c^2}{\omega_b^3\, c_b^2}  ~. 
\label{e5}
\end{equation} 
The pump pulse envelope, $a$, is further normalized such that the average square of the electron quiver velocity in the pump laser field, measured in units of $c^2$, is $|a|^2$, so that
\begin{equation}\displaystyle
\overline{v_{ea}^2}=c^2|a|^2.
\label{e6}
\end{equation}
Then, the average square of the electron quiver velocity in the seed laser field and in the Langmuir wave field are given by 
\begin{equation}\displaystyle
\overline{v_{eb}^2}=c^2|b|^2 \frac{\omega_a}{\omega_b} , \;\quad
\overline{v_{ef}^2}=c^2|f|^2 \frac{\omega_a}{\omega_f}.
\label{e7}
\end{equation}
The  3-wave coupling constant can be written as \cite{Kruer1988}
\begin{equation}\displaystyle
V_3=k_f c\sqrt{\frac{\omega_e}{8\omega_b}}\, 
\label{e8}
\end{equation}
where $k_f$ is the wave number of the resonant Langmuir wave
\begin{equation}\displaystyle
k_f=k_a+k_b,\;\; k_ac=\sqrt{\omega_a^2-\omega_e^2},\;\; k_bc=\sqrt{\omega_b^2-\omega_e^2}~.
\label{e9}
\end{equation}
The frequency resonance condition is 
\begin{equation}\displaystyle
\omega_b+\omega_f=\omega_a\, ,
\label{e10}
\end{equation}
where $\omega_f\approx \omega_e$ is the Langmuir wave frequency in a not too hot plasma. 
The nonlinear frequency shift coefficient $R$ can then be put as \cite{Litvak1969,Max1974,Sun1987}
\begin{equation}\displaystyle
R=\frac{\omega_e^2\omega_a}{ 4\omega_b^2}  .
\label{e11}
\end{equation}

This hydrodynamic model is applicable for the pump pulse intensity $I_0$ smaller than that at the threshold of the resonant Langmuir wave breaking $I_{\rm br}$. The motivation for studying specifically such regimes is that for deep wavebreaking regimes the BRA efficiency is lower \cite{Malkin_99_PRL,Malkin_00_POP}.

\section{Universal variables}
The above equations will be solved for a small Gaussian initial seed and constant initial pump with a sharp front.
After entering the pump depletion stage,  the leading amplified spike (propagating directly behind the seed pulse) grows and contracts (since it depletes the pump faster and faster, as it grows).    Thus the spike  becomes of much shorter duration than the elapsed amplification time,  attaining  the universal features of a classical $\pi$-pulse before the REN becomes important. This prepares universal initial conditions for entering the REN regime.
To expose this universality, it is helpful to change $z$ and $t$ variables to dimensionless variables 
\begin{eqnarray}
&\displaystyle\tau=\left(1+\frac{c_a}{c_b}\right)^{1/3}R^{1/3} V_3^{2/3}a_0^{4/3}\, \frac{L-z}{c_b}~,  \label{e12}\\
&\displaystyle\zeta= \left(1+\frac{c_a}{c_b}\right)^{-1/3}
R^{-1/3} V_3^{4/3} a_0^{2/3} \left(t-\frac{L-z}{c_b}\right), \nonumber
\end{eqnarray}
where $\tau$ measures the elapsed amplification time (or the distance traversed by the original seed front), $\zeta$ measures the distance (or delay time)  from the original seed  front; $L$ is the plasma width  and $a_0$ is the input pump amplitude; the seed is injected into the plasma at $z=L$, $t=0$ and  meets immediately the pump front injected into the plasma at $z=0$, $t=-L/c_a$.

Then, defining new wave amplitudes $a_1$, $f_1$ and $b_1$ by formulas 
\begin{eqnarray}
&\displaystyle a=a_0 a_1~, \label{e13} \\
&\displaystyle f= -a_0 \left(1+\frac{c_a}{c_b}\right)^{1/2} f_1 ~,\label{e14}\\
&\displaystyle b=\left(\frac{V_3a_0^2}{R}\right)^{1/3} \left(1+\frac{c_a}{c_b}\right)^{1/6} b_1~,
\label{e15}
\end{eqnarray}
and neglecting the ``slow" time derivative of the pump amplitude compared to the ``fast" time derivative of the pump amplitude, one obtains the following universal equations containing just one parameter $Q$:
\begin{eqnarray}
&\displaystyle a_{1\zeta}=-b_1 f_1\, ,\label{e16}\\
&f_{1\zeta}=a_1 b_1^*,\label{e17}\\
&b_{1\tau}=a_1 f_1^*-\imath Q b_{1\zeta\zeta}+
\imath |b_1|^2 b_1,\label{e18}\\
&\displaystyle Q=\frac{(k_a+k_b)^2c^2\omega_b c_b'}{4\omega_e\omega_a(c_a+c_b)}.\label{e19}
\end{eqnarray}
The  parameter {\it Q} characterizes the group velocity dispersion of amplified pulse and depends only on the ratio of the plasma to laser frequency $q \equiv\omega_e/\omega_b$.  In strongly under-critical plasmas, where $q\ll 1$, one has $Q=q/2$;  in nearly critical plasmas, where $q\rightarrow 1$, one has $Q=0.5/\sqrt{1-q^2}\gg 1$. 
  
In strongly undercritical plasmas, which is of major interest here, the amplified pulse intensity $I$, fluence $w$ and effective duration $\Delta t_b$ can then be expressed in these variables as 
\begin{eqnarray}
&\displaystyle I=\frac{G|b_1|^2\omega_e}{4\lambda_b}\left(\frac{I_0^2}{2I_{\rm br}^2}\right)^{1/3}, \label{e20}\\
&\displaystyle w=\frac{G\tau}{\lambda_b}\left(\frac{I_0}{4I_{\rm br}}\right)^{1/3}, \label{e21}\\
&\displaystyle \Delta t_b=\frac{w}{\max_\zeta I}=\frac{4\tau}{\omega_e \max_\zeta |b_1|^2}\left(\frac{I_{\rm br}}{2I_0}\right)^{1/3},
\label{e22}
\end{eqnarray}
where 
\begin{equation}\displaystyle
G=m_e^2c^4/e^2=0.3 \,\rm J/cm,  \;\;\;\lambda_b=2\pi/k_b\, ,
\label{e23}
\end{equation}
and 
\begin{equation}\displaystyle
I_{\rm br}=n_em_ec^3q/16
\label{e24}
\end{equation} 
is the threshold pump intensity for resonant Langmuir wave breaking.  The formula for fluence here assumes nearly complete pump depletion.

Eqs.~(\ref{e16})- (\ref{e18}) will be solved now for small input Gaussian seed pulses of the form $$\displaystyle b_1(\zeta,0)= \frac{b_{1_0}}{\sqrt{D\pi}}
\exp\left[-\frac{(\zeta-\zeta_{0})^2}{D}\right]$$ 
with $b_{1_0}=0.05$, $D=1$ and $\zeta_{0}=10$. 
No auxiliary chirping of the seed pulse is needed here, though it may be useful in less undercritical plasmas \cite{Toroker_12_PRL}.

\section{Dispersionless REN regime}

First, consider extremely undercritical plasmas where the group velocity dispersion can be neglected, so that the approximation $Q=0$ is good enough. 

\begin{figure}[ht]
\includegraphics[width=0.5 \textwidth]{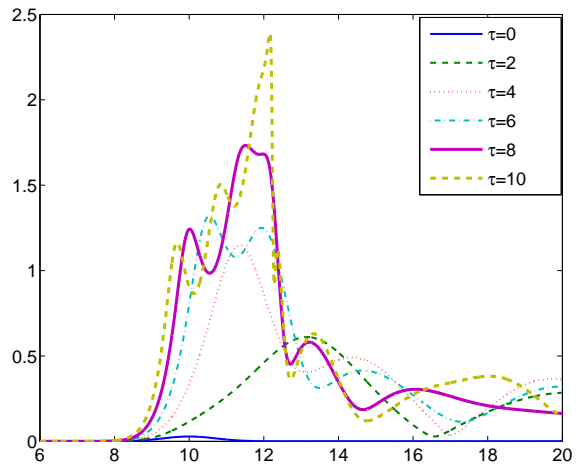}
\caption{The amplified pulse amplitude $|b_1|$ vs. the delay time $\zeta$ at several values of the amplification time $\tau$ in extremely undercritical plasma. }
\label{f1} \end{figure}

\begin{figure}[ht]
\includegraphics[width=0.5 \textwidth]{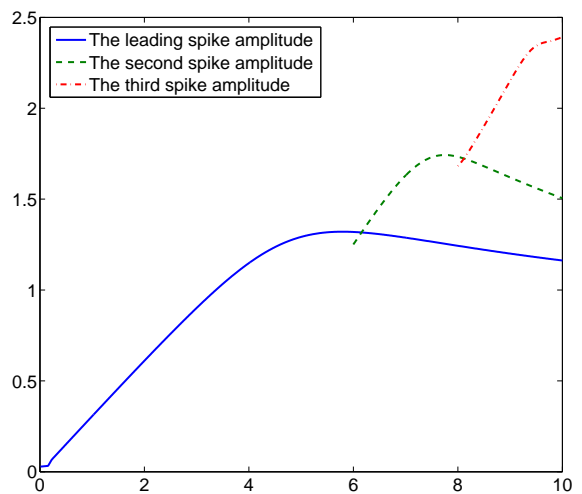}
\caption{The maximal amplitude of the amplified  pulse $\max_\zeta |b_1|$ as a function of the amplification time $\tau$ in extremely undercritical plasma.} 
\label{f2} \end{figure}

Fig.~\ref{f1} shows the rescaled amplified pulse amplitude $|b_1|$ as a function of the delay  time $\zeta$ at several amplification times $\tau$ for $Q=0$.  
The amplified pulse may have its maximum amplitude from either the first spike or from later spikes. This maximum is  depicted in   Fig.~\ref{f2}. The initial nearly linear part of the curve  in   Fig.~\ref{f2} corresponds to the classical $\pi$-pulse regime.
In the REN regime, the leading spike growth saturates, while the second spike  grows, reaching even higher intensity. 
Then the second spike growth saturates, while the third spike  grows, reaching even higher intensity yet. The spikes do not filament and remain distinguishable for a while.  As seen from the Fig.~\ref{f2}, the top amplified pulse amplitude can be nearly double  the largest leading spike amplitude, so that output intensity can be nearly 4 times the leading spike theoretical limit.
 

\section{The effect of group velocity dispersion}

\begin{figure*}[ht]\nonumber
\hskip-1cm\includegraphics[width=0.38 \textwidth]{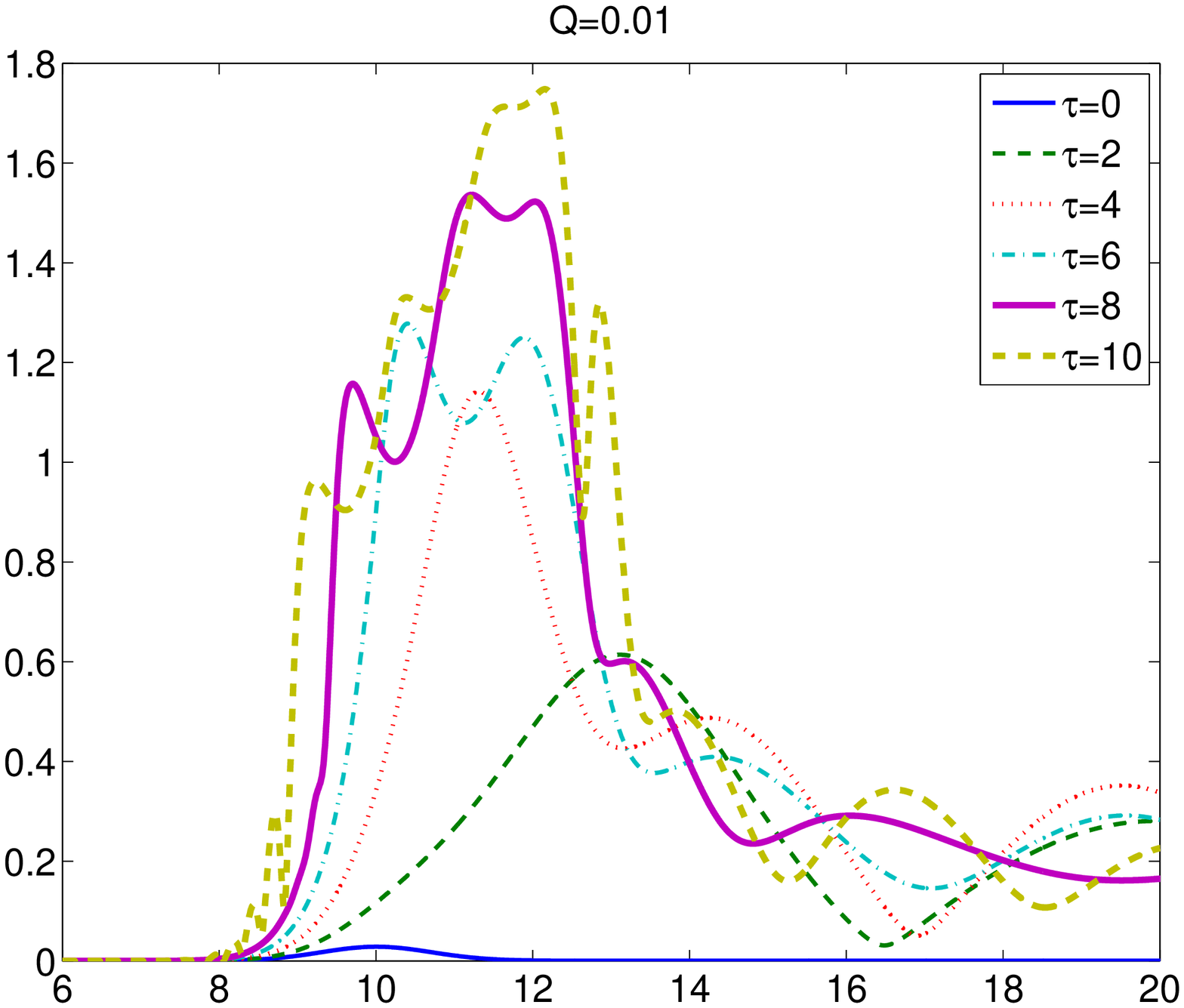}
\hskip-1cm\includegraphics[width=0.38 \textwidth]{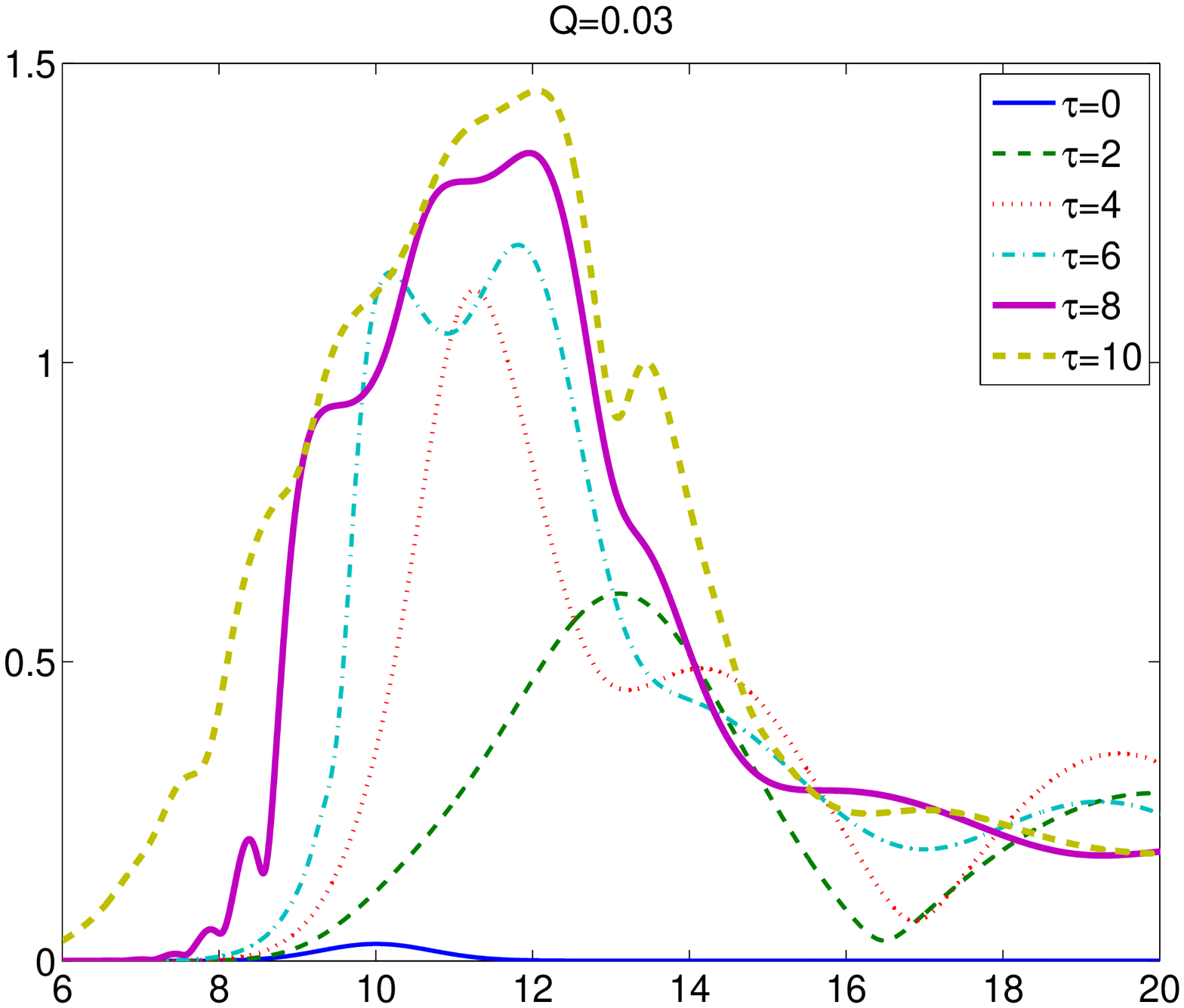}
\hskip-1cm\includegraphics[width=0.38 \textwidth]{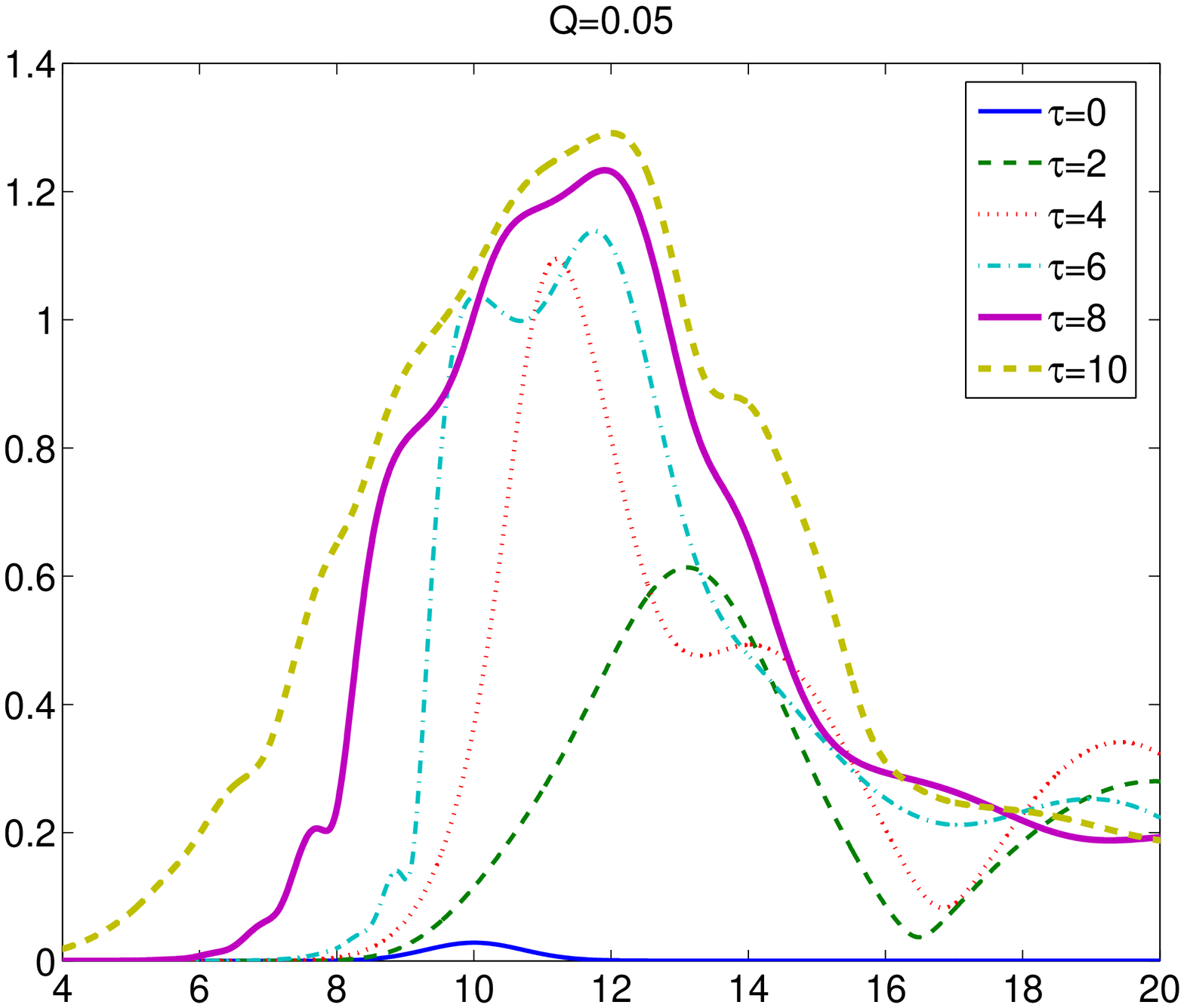}
\caption{The amplified pulse amplitude $|b_1|$ vs. the delay time $\zeta$ at several values of the amplification time $\tau$ and the dispersion parameter $Q$.}
\label{f3} 
\end{figure*}

\begin{figure}[ht]
\includegraphics[width=0.4 \textwidth]{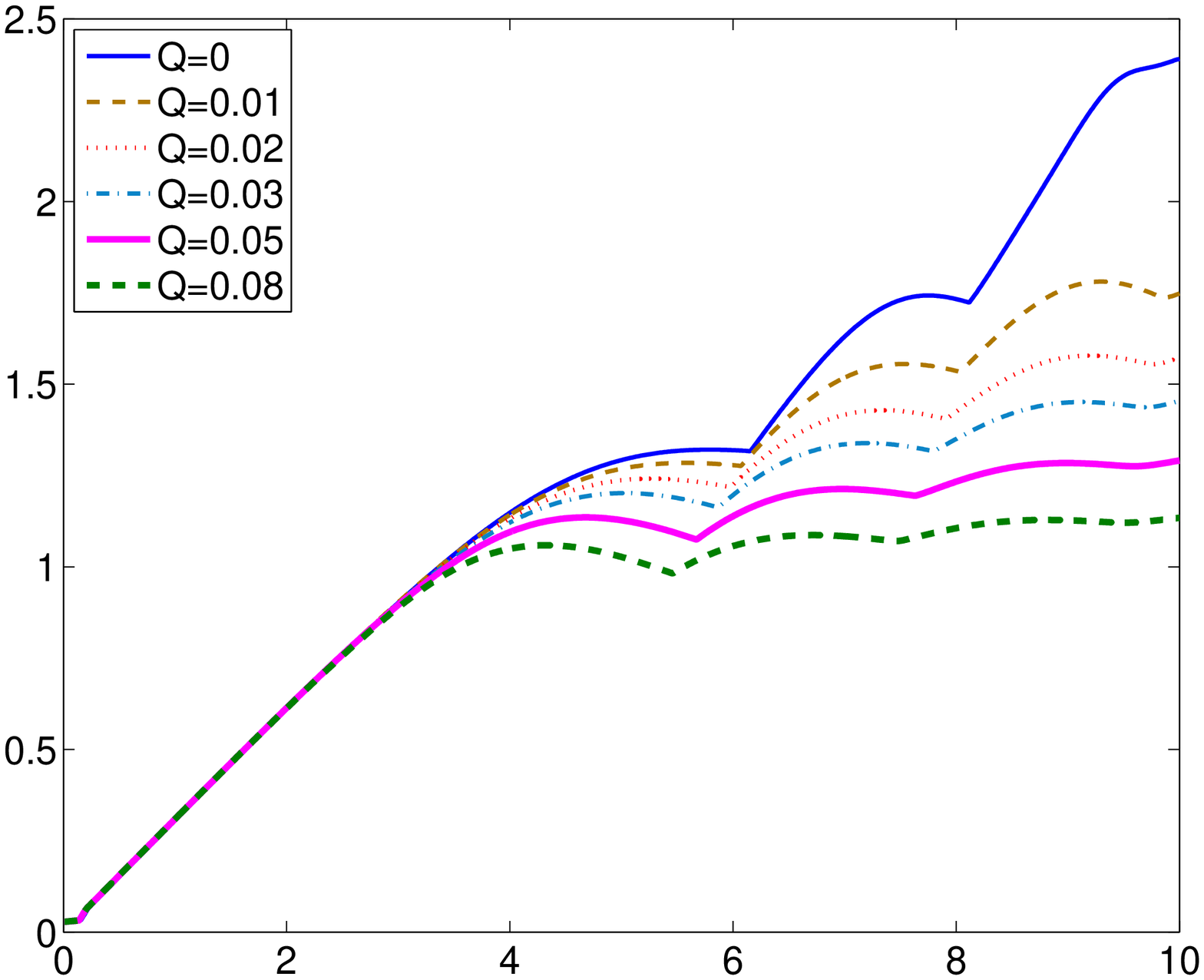}
\includegraphics[width=0.4 \textwidth]{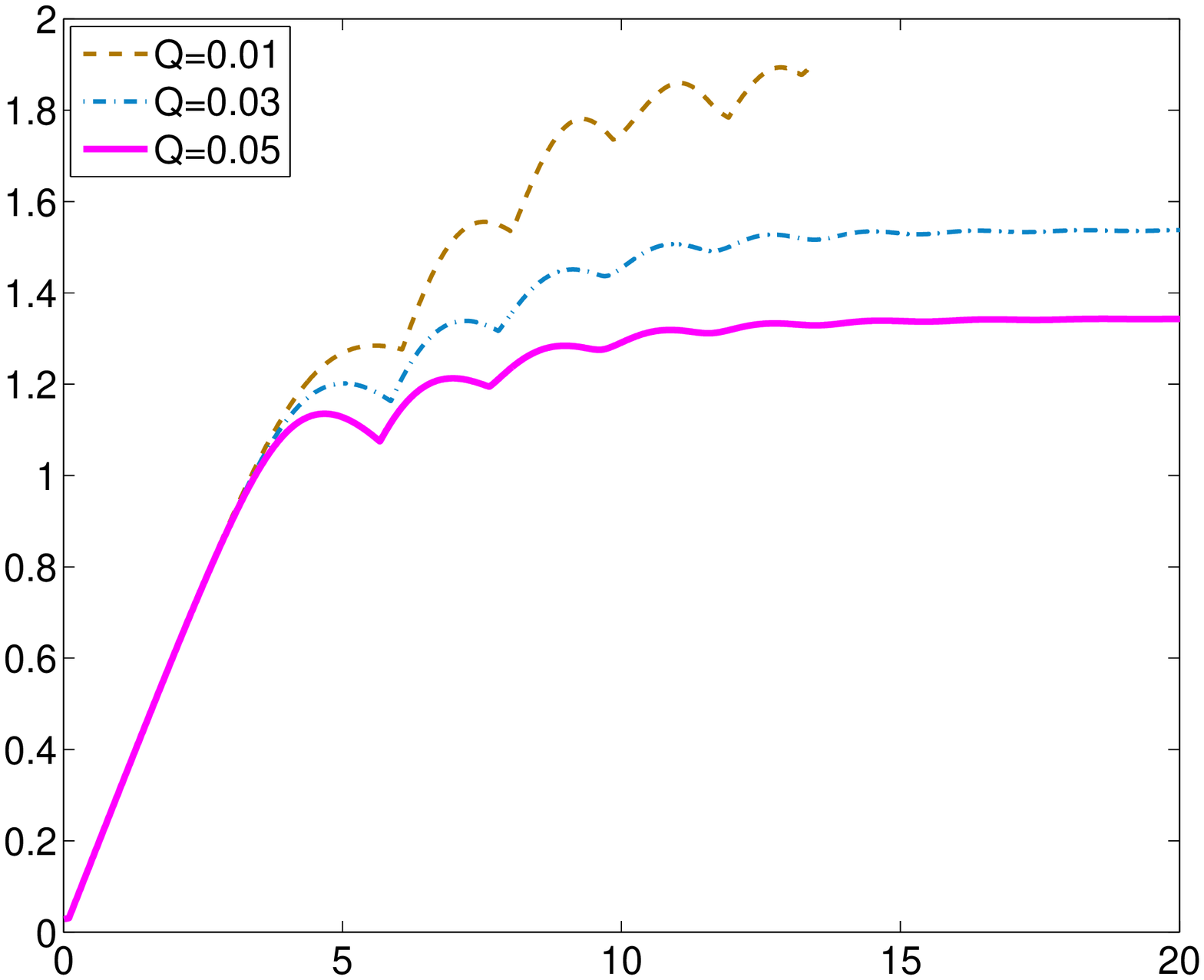}  
\caption{The maximal amplitude of the amplified  pulse $\max_\zeta |b_1|$ as a function of the amplification time $\tau$ for several values of the dispersion parameter $Q$.}
\label{f4} 
\end{figure}

For less extreme, though still strongly undercritical plasmas, the group velocity dispersion can become important in the REN regime. This is in contrast to the $\pi$-pulse regime for which the group velocity dispersion is negligible in strongly undercritical plasmas  \cite{Malkin_07_PRL}.  The importance  of even  rather small group velocity dispersion  in the REN regime is illustrated in Figs.~\ref{f3} and \ref{f4} which show the dispersion effect at small $Q\approx q/2$.

Fig.~\ref{f4} shows  the maximum pulse amplitude $\max_\zeta |b_1|$ as a function of the amplification time $\tau$.  
Note that  the $\pi$-pulse regime corresponds to the joint straight part of the curves located approximately at times $\tau < 2$.  
Here,  there is indeed no {\it Q}-dependence, indicating the negligibility of the group velocity dispersion.  
However, in the REN regime ($\tau > 3$), the {\it Q}-dependence becomes increasingly prominent. 
Larger $Q$ corresponds to smaller pulse amplitudes, because group velocity dispersion tends to stretch the pulses.  
It also tends to delay the onset of the longitudinal filamentation instability, thus enabling yet larger output fluences if not intensities.

It can be seen from Figs.~\ref{f3} and \ref{f4}  that  significant additional growth of the amplified pulse intensity and fluence can occur not only beyond the classical $\pi$-pulse regime, but even after the leading spike saturation. 
In extremely undercritical plasmas, where $Q\lesssim 0.01$, subsequent to the  leading spike saturation, the amplified pulse intensity and  fluence can increase further by a factor of about  3. 
In denser, but still strongly undercritical plasmas with $Q\sim 0.02 - 0.03$, the amplified pulse growth subsequent to the  leading spike saturation can be  about  2-fold in intensity and about  4-fold in fluence. 

For example, for  $\lambda_b=1/4\,\mu$m and $I_0=I_{\rm br}/2$, and $Q=0.025$ (corresponding to $\omega_e/\omega_b=0.05$), the fluence achievable in the REN regime is 120 kJ/cm$^2$. 
Here, the plasma concentration is $n_e=4.5\times 10^{19}\,\rm cm^{-3}$ and  the input pump intensity is $I_0=1.7\times 10^{14}\,\rm W/cm^2$; the pump duration is 0.7 ns, the amplified pulse output duration is 94 fs and the intensity is $1.2\times 10^{18}\,\rm W/cm^2$. This is more than 10 times larger than intensities reached in the recent numerical simulations \cite{Trines_11_PRL}.  In these simulations, the REN
regime was not reached apparently because of instabilities.  These instabilities arise from noise, which in particle-in-cell codes might even exceed real plasma noise. In any event, the instabilities might have been suppressed, say, by applying selective detuning techniques \cite{Malkin_00_PRL,Malkin_00_POP,Tsidulko_00_PRL,Solodov_04_PRE,Malkin_14-EPJST} (which were not employed in \cite{Trines_11_PRL}).  The same applies to the paper \cite{Trines_11}
where somewhat larger output intensities, like $4\times 10^{17}\,\rm W/cm^2$ , were reported numerically for much larger pump intensities  $I_0\approx 30I_{\rm br} \gg I_{\rm br}$ not discussed here.
Note that the ability to compress laser pulses from ns to 100 fs duration may allow direct BRA of currently available powerful 1/4 micron wavelength ns laser pulses to ultrahigh powers.
The regimes found here can further enhance multi-step BRA schemes \cite{Fisch_03_POP, Malkin_05_POP}, as well as possible combinations of such schemes with other currently considered methods of producing ultra-high laser intensities, like \cite{Shvets_98_PRL,PRL-2010-Lancia,UFN-2011-Korzhimanov,RevModPhys-2012-Piazza, Mourou_2012,UFN-2013-Bulanov,PRL-2013-Weber,PRL-2014-Tamburini}.

\section{Summary}
In summary,  an amplification regime is identified here  wherein output pulse intensities and fluences substantially surpass the previous theoretical limit for strongly undercritical plasmas.
The new intensity and fluence limits are produced
by the initially sub-dominant spikes of the amplified wavetrain, which were not 
previously thought to be important for achieving the largest output pulses.
In addition, the amplified pulse regular group velocity dispersion, in spite of being small in  strongly undercritical plasmas, is shown nevertheless to be capable of a further delaying the pulse filamentation, thus allowing a further pulse amplification to even larger output fluences.

\section{Acknowledgments}
This work was supported by DTRA 
HDTRA1-11-1-0037, by NSF PHY-1202162, and by
the NNSA SSAA Program under Grant No~ DE274-FG52-08NA28553.

\providecommand{\noopsort}[1]{}\providecommand{\singleletter}[1]{#1}%

\end{document}